%% file: beuther_porto2004.tex
\documentclass{kapproc} % Computer Modern font calls
\usepackage{procps} 
\usepackage[dvips]{graphicx}
\upperandlowercase

\kluwerbib % will produce this kind of bibliography entry:

\begin{document}
\input{refs.tex}

%%% Page limits: minimum 6 and maximum 14 (for a 45 minutes talk)

\articletitle{Precursors of UCH{\sc ii} regions \&~ \\ the 
evolution of massive outflows}

\begin{minipage}[t]{6.5cm}
\begin{flushleft}
\author{Henrik Beuther}
\affil{Harvard-Smithsonian Center for Astrophysics\\
60 Garden Street\\ Cambridge, MA 02138 \\ USA}
\email{hbeuther@cfa.harvard.edu}
\end{flushleft}
\end{minipage}
\begin{minipage}[t]{6.5cm}
\begin{flushleft}
\author{Debra Shepherd}
\affil{National Radio Astronomy Observatory\\ 
Socorro, NM 87801 \\ USA}
\email{dshepher@nrao.edu}
\end{flushleft}
\end{minipage}

\begin{abstract} Since this contributions was meant to cover two
subjects which are both in the field of massive star formation
but which in its details can be discussed separately, this paper is
divided in two sections. First, we present characteristics of
precursors of UCH{\sc ii} regions and their likely evolutionary
properties.  The second section discusses massive molecular outflows,
their implications for high-mass star formation, and a possible
evolutionary sequence for massive outflows.  \end{abstract}


\section{Precursors of UCH{\sc ii} regions}
\label{precursors}

Evolutionary scenarios suggest that High-Mass Starless Cores (HMSCs)
represent the earliest evolutionary stage of massive star formation
(e.g., \cite{evans2002}). The next observable evolutionary stage is
when a High-Mass Protostellar Object (HMPO) forms in the core
producing strong millimeter continuum and mid-infrared emission but no
detectable centimeter continuum due to free-free emission from ionized
gas (e.g., \cite{molinari1996,sridha}).  The lack of detectable
free-free emission may be because the protostar has not reached the
main sequence or because the accretion rate onto the stellar surface
is high enough to quench a developing HII region (e.g.,
\cite{walmsley1995, churchwell2002}). Hot Cores are considered to
belong to the HMPO stage. Soon after the hot core is formed, the
central massive protostar produces adequate ionizing radiation to form
a hypercompact, unresolved, most likely optically thick H{\sc ii}
region (HCH{\sc ii}). The HCH{\sc ii} region may still be partially
quenched by infalling gas which can hinder the expansion of the
hypercompact H{\sc ii} region (e.g.,
\cite{walmsley1995,keto2003}). Eventually, the hypercompact H{\sc ii}
region begins to expand forming the well-studied ultracompact H{\sc
ii} region (e.g., \cite{wc89,kurtz1994}) and later a more evolved
H{\sc ii} region. Theoretical model calculations estimate an
evolutionary time-scale for massive star formation of about
$10^5$\,yrs, assuming that massive stars form via accretion-based
processes similar to low-mass star formation (\cite{mckee2002}).  A
key difference between low- and high-mass star formation is that
massive stars reach the main sequence while still actively accreting;
low-mass stars end their accretion phase before they reach the main
sequence. Here, we discuss the properties of the precursors of
ultracompact H{\sc ii} regions, namely HMPOs and HMSCs.

\subsection{High-Mass Protostellar Objects (HMPOs)}

There are several review articles that discuss the earliest
stages of massive star formation and cover different aspects of the
topic in more depth than possible here (e.g.,
\cite{garay1999,kylafis1999,kurtz2000,churchwell2002,cesaroni2004}).
Therefore, we focus on the basic characteristics of HMPOs and
discuss a few recent results in this field.

\subsubsection{Basic characteristics}

The far-infrared colors of HMPOs are similar to those of UCH{\sc ii}
regions (e.g., \cite{wc89,sridha}). This is mainly due to the fact
that they resemble each other with regard to their gas/dust
temperatures and the presence of dense gas. A major observational
difference is that HMPOs show no or only weak cm continuum emission,
thus the central objects have not yet produced a significant ionized
region to trigger the necessary free-free emission. While most of the
UCH{\sc ii} region luminosity is due to a central H-burning
star, a large fraction of the HMPO luminosity is still expected to be
due to accretion (e.g., \cite{sridha}). Furthermore, the average
observed NH$_3$(1,1) linewidth toward HMPOs is $\Delta v \sim
2.1$\,km/s compared to average values toward UCH{\sc ii} regions of
$\Delta v \sim 3.1$ (\cite{sridha}). This difference implies less
turbulent motions at the HMPO stage.

Furthermore, often H$_2$O and/or Class {\sc ii} CH$_3$OH maser
emission is observed toward HMPOs (e.g.,
\cite{walsh1998,beuther2002c,codella2004}). The existence of one or
the other type of maser emission is widely regarded as a good signpost
for massive star formation (in spite of H$_2$O maser emission being
also observed toward low-mass star-forming regions), but nevertheless,
there is no general agreement about the physical processes generating
the maser emission.  There is indicative evidence that both maser
types are produced either within accretion disks or molecular outflows
(e.g.,
\cite{norris1998,walsh1998,torrelles1998,minier2000,codella2004}).
Although H$_2$O and Class {\sc ii} CH$_3$OH masers are produced during
the early stages of massive star formation, it appears that in general
they cannot be used to derive the underlying physical processes of the
regions.  This statement does not exclude the possibility that in
selected sources very high-spatial resolution and proper motion
studies of masers can constrain the processes triggering the maser
emission (for prominent recent examples see, e.g.,
\cite{torrelles2003,pestalozzi2004}).

\subsubsection{Millimeter continuum emission}

Observations of HMPOs in (sub)mm continuum emission have shown that
they exhibit strong dust continuum emission from massive dust and gas
cores (e.g.,
\cite{molinari2000,beuther2002a,mueller2002,williams2004}). Even with
the rather coarse spatial resolution of most single-dish
investigations ($\geq11''$ beam), the gas cores often show filamentary
and multiple structures. The density distribution of the cores are
usually well fitted by power-law distributions and the power-law
indices are in most cases around 1.6 (e.g.,
\cite{beuther2002a,mueller2002}), similar to the power-law indices of
young low-mass cores (e.g., \cite{motte1998}). Furthermore,
multi-wavelength investigations show that the dust opacity index
$\beta$ toward HMPOs is significantly lower than toward UCH{\sc ii}
regions (mean values of 0.9 versus 2.0,
\cite{williams2004,hunter1997}). In low-mass cores, low dust opacity
indices are attributed either to increasing opacity or grain growth at
early evolutionary stages (e.g.,
\cite{hogerheijde2000,beckwith2000}). In contrast, Williams et
al.~(2004) show for their sample that the lowest observed values for
$\beta$ are associated with lower rather than higher opacities. Thus,
these observations are consistent with grain growth in the early
evolutionary stages of the dense cores. However, a point of caution
remains because significant grain growth is expected to take at least
$10^5$ years (e.g., \cite{ossenkopf1994}) whereas the whole massive
star formation process might only last for $\sim 10^5$ years
(\cite{mckee2002}). If these time-scales are correct it may be
difficult to produce similarly large grains in high-mass cores
compared to their low-mass counterparts and the observed low
values of $\beta$ might require a different interpretation.

\subsubsection{Fragmentation and the Initial Mass Function (IMF)}

One important question for the whole cluster formation process is: At
what evolutionary stage does the Initial Mass Function gets
established?  Do the initial fragmentation processes of the massive
cores already show signatures of the IMF, or is competitive accretion
during the ongoing cluster formation the main driver of the IMF (e.g.,
\cite{bonnell2004})?  Within the last two years, there have been
published a few very different studies which investigate the
protocluster mass distribution and its connection to the IMF:
\cite{shirley2003}, analyze single-dish CS observations toward a
sample of 63 massive H$_2$O maser sources with $\sim 30''$ resolution,
\cite{williams2004}, do a similar study based on single-dish (sub)mm
continuum observations with $8''-15''$ resolution, and
\cite{beuther2004c}, observe one HMPO in the mm continuum emission at
$1.5''\times 1''$ resolution with the Plateau de Bure
Interferometer. The main difference between the two single-dish
studies and the interferometric investigation is that the single-dish
observations average out whole young clusters and thus need many
clusters to derive a cumulative mass distribution of the sample of
protoclusters, whereas Beuther \& Schilke (2004) only investigate one
young cluster but are capable of resolving many sources within this
cluster.  All of the above mentioned studies preferentially select
regions that are in the earliest evolutionary stages of star formation
and most have not yet formed detectable UCH{\sc ii} regions.  Figure
\ref{imf} summarizes the results from the three studies.

\begin{figure}
\caption{\footnotesize Compilation of mass distributions from very
young high-mass star-forming regions: the cumulative mass distribution
to the left is from single-dish CS observations of a sample of massive
H$_2$O maser sources (\cite{shirley2003}, LS and RE correspond to
Least Square and Robust estimation fits). The cumulative mass spectrum
in the middle is from single-dish mm dust continuum observations of a
sample of HMPOs (\cite{williams2004}, the dotted and full lines
correspond to far and near kinematic distances, respectively). The
protocluster mass function to the right is from high-spatial
resolution PdBI observations of one HMPO IRAS\,19410+2336
(\cite{beuther2004c}, the full line is the best fit to the data, and
the dashed and dotted lines present the IMFs derived from
\cite{salpeter1955}, and \cite{scalo1998}). }
\label{imf}
\end{figure}

Since the observations and the analysis are very different, the
completeness limits of the Shirley et al.~and Williams et al.~studies
are different as well ($>1000$ and $>10$\,M$_{\odot}$,
respectively). Nevertheless, both studies find above their
completeness limits power-law distributions comparable to the
IMF. Similarly but on much smaller spatial scales, Beuther \& Schilke
(2004) also find a good correspondence between the protocluster mass
spectrum in IRAS\,19410+2336 compared with the IMFs as derived by,
e.g, \cite{salpeter1955,scalo1998}. These independently derived
results are encouraging support for the idea that the early
fragmentation of massive star-forming cores is responsible for the
determination of the IMF, and not processes which might take place at
evolutionary later stages like competitive accretion. An
additional result from the single-dish observations alone is that the
star-formation efficiency for the sample of protoclusters is
relatively mass invariant (\cite{williams2004}).

\subsubsection{Chemistry}

Young massive star-forming regions and especially Hot Cores are known
to have an incredibly rich chemistry (e.g., the rich spectra toward
the Orion-KL Hot Core, e.g., \cite{blake1987,schilke1997b}). Here,
we present recent submillimeter observations of the prototypical Hot
Core Orion-KL.  A more detailed discussion of core chemistry is
presented in Paola Caselli's contribution to this volume.

\cite{beuther2004g}, used the Submillimeter Array (SMA) to image the
submm line and continuum emission of Orion-KL at sub-arcsecond
resolution.  In the submm continuum, the enigmatic source ``I'' could
be disentangled from the Hot Core, and source ``n'' was detected for
the first time shortward of 7\,mm (\cite{beuther2004g}). Additionally,
within one spectral setup of the SMA (4\,GHz combining the upper and
lower sideband) approximately 150 molecular lines were observed
(Beuther et al.~in prep.).  Figure \ref{orion} shows example line
images which highlight the spatial variations observed between the
different molecular species: SiO is detected mainly from the outflow
emanating from source ``I'', but all other molecules peak offset from
source ``I''.
\begin{figure}[h]
\begin{center}
\caption{\footnotesize Submm line images obtained with the SMA at
865\,$\mu$m (Beuther et al.~in prep.). The grey-scale and contours
show the molecular emission as labeled at the bottom left of each
panel. The three stars mark the positions of source ``I'', the Hot
Core mm continuum peak, and source ``n'' (labeled at the bottom-right).
}
\end{center}
\label{orion}
\end{figure}
Typical hot core molecules like CH$_3$CN or CH$_3$CH$_2$CN show the
hot core morphology previously reported from, e.g., NH$_3$
observations (\cite{wilson2000}).  Contrary to this, CH$_3$OH does
show some hot core emission as well, but there is an additional strong
CH$_3$OH peak further south associated with the so called compact
ridge emission. Other molecules like C$^{34}$S or C$_2$H$_5$OH show
even different spatial emission, tracing partly the hot core and the
compact ridge but also showing an additional peak to the north-west
that is spatially associated with IRC6. We still do not properly
understand the underlying reasons for this spatial molecular
diversity, but these observations of the closest Hot Core, Orion-KL,
stress that we must be cautious in our interpretations of spatial
structures for other Hot Cores that are typically at much greater
distances.

\subsection{High-Mass Starless Cores (HMSCs)}

Prior to the infrared satellites ISO and MSX, there was no
observational evidence for High-Mass Starless Cores, and HMPOs were
considered to be the youngest observationally accessible stage of
massive star formation. The detection of Infrared Dark clouds (IRdCs)
and their association with mm dust continuum emission changed that
situation (e.g., \cite{egan1998}), and since then observational
studies of these HMSCs/IRdCs have begun to shed first light on
this evolutionary stage. \cite{evans2002}, outlined potential
properties of HMSC, and recently \cite{garay2004}, presented an
analysis of four such sources. Here, we present an approach to
identify and study HMSCs based on a comparison of previously
derived mm continuum maps of HMPOs (\cite{beuther2002a}) with the MSX
database (Sridharan et al.~in prep.).

\begin{figure}[htb]
\begin{center}
\includegraphics[angle=-90,width=8cm]{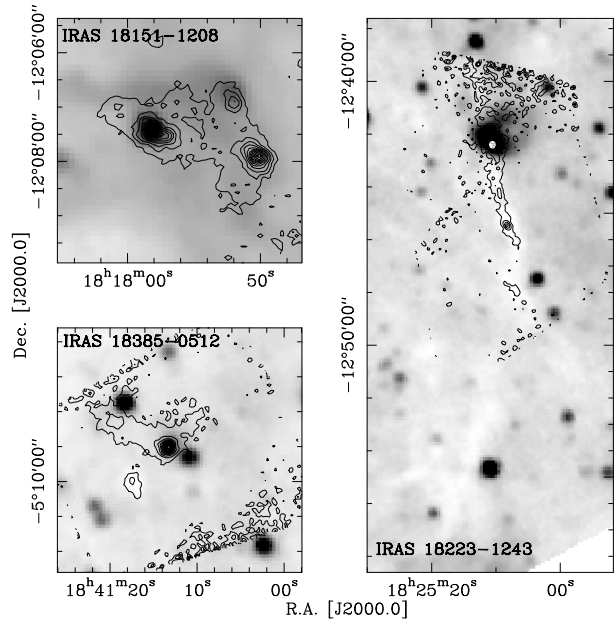}
\end{center}
\caption{\footnotesize The grey-scale shows the 8\,$\mu$m emission 
from the MSX satellite, and the contours present the 1.2\,mm continuum
emission first presented in \cite{beuther2002a}.}
\label{hmsc}
\end{figure}

Based on an initial study of HMPOs associated with IRAS sources
(\cite{sridha}), \cite{beuther2002a}, observed serendipitously within
the same fields of view a number of secondary mm-peaks which were not
associated with IRAS sources.  Correlation of these mm-peaks with
the MSX mid-infrared observations revealed a number of mm-peaks which
are not only weak in the mid-infrared but which are seen as absorption
shadows against the galactic background (Fig.~\ref{hmsc}, Sridharan et
al.~in prep.).

Observing these HMSCs with the Effelsberg 100\,m telescope in NH$_3$
(Sridharan et al.~in prep.), the derived average temperatures are
around 15\,K compared to the average temperatures of the more evolved
HMPOs of 22\,K.  Furthermore, we find on average a smaller linewidth
of 1.6\,km/s toward the HMSCs in comparison to a mean value of
2.1\,km/s observed toward the HMPOs. With gas masses of the order a
few hundred M$_{\odot}$, no mid-infrared emission, cold temperatures
and less turbulence implied by the small line-widths, these sources are
promising candidates for being HMSCs on the verge of forming
massive clusters. Ongoing  single-dish and interferometric
follow-up observations of this and other samples will hopefully
improve our understanding of the initial conditions of massive star
formation significantly.

\section{Massive molecular outflows: an evolutionary scenario}

In contrast to low-mass outflows, there is no general census for the
understanding of massive molecular outflows.  For recent reviews on
different aspects of massive molecular outflows see
\cite{bachiller1999,churchwell1999,richer2000,shu2000,koenigl2000,shepherd2003b,beuther2004f}.

Two competing scenarios have been put forth to explain the formation
of massive stars: one possibility is to extend the low-mass models to
high-mass star formation and form massive stars via enhanced accretion
processes through accretion disks (e.g.,
\cite{jijina1996,norberg2000,yorke2002,mckee2003}).  Contrary to this,
at the dense centers of evolving massive clusters, competitive
accretion could become dominant and it might even be possible that
(proto)stellar mergers occur and form the most massive objects via
coagulation processes (e.g.,
\cite{bonnell1998,stahler2000,bally2002,bonnell2004}).  Massive
molecular outflows can help to differentiate between these scenarios:
while the enhanced accretion model predicts collimated molecular
outflows with properties similar to those for low-mass outflows, it is
unlikely that highly-collimated structures could exist in the
coalescence scenario and the outflow energetics would be significantly
different.

\subsection{Summary of observational constraints}

Observations of massive molecular outflows are unfortunately not
conclusive.  On the one hand, some observations find that molecular
outflows from high-mass star-forming regions appear less collimated
than outflows from low-mass regions (e.g.,
\cite{shepherd1996b,ridge2001,wu2004}).  However, it has been shown
that the lower collimation factors of massive outflows observed with
single-dish telescopes could simply be due to the on average larger
distances of high-mass star-forming regions (\cite{beuther2002b}).  A
few interferometric observations with $0.1''-10''$ resolution suggest
that outflow opening angles from at least some young, early-B stars
can be significantly wider than those seen toward low-mass protostars
(e.g., \cite{shepherd1998,shepherd2001,shepherd2003}).  Thus, greater
distances and the associated lower angular resolution is not the only
reason for the observed poor collimation in some massive outflows.
There is also evidence that the energetics in at least one massive
outflow from an early B star in W75N are different to those in
low-mass outflows, although exactly what this implies is unclear
(\cite{shepherd2003}).

In recent years a few examples of collimated massive outflows from
early B stars have been found (e.g.,
\cite{beuther2002d,gibb2003,su2004}).  A comparison of
position-velocity diagrams from collimated massive molecular outflows
with their low-mass counterparts reveal similar signatures from
outflows of all masses (\cite{beuther2004d}).  Moreover, a
near-infrared analysis of a HMPO found collimated outflows with
properties appearing to be scaled-up versions of low-mass outflows
(\cite{davis2004}).  This latter group of observations favors an
accretion scenario for the formation of massive stars that is a scaled
up version of the low-mass scenario.

For both well-collimated and poorly collimated outflows, relations
between outflow and source parameters hold over many orders of
magnitude (e.g., \cite{richer2000,beuther2002b,wu2004}).  Richer et
al.~point out that these correlations do not necessarily argue for a
common entrainment or driving mechanism for sources of all
luminosities given that one would expect that most physically
reasonable outflow mechanisms would generate more powerful winds if
the source mass and luminosity are increased. It is interesting to
note that no well-collimated outflows have been observed toward a
(proto)-O star (e.g., \cite{shepherd2003b}, and references therein,
\cite{sollins2004b}).  Thus, the link between accretion and outflow is
not as well established for forming O stars as for early B stars and
formation by coalescence may still be a possibility. Table
\ref{summary} presents a number of relevant references for
observations which shed light on massive outflows.

\begin{table}[htb]
\caption{Summary of outflow results.  This table does not claim
completeness but gives an overview of relevant publications
connected to the question of outflow evolution. \label{literature}}
\label{summary}
\begin{tabular}{lcr}
\hline \hline
Source & Results & Ref \\
\hline
\multicolumn{3}{c}{Low-spatial resolution single-dish studies}\\
\hline
122 UCH{\sc ii}s & ubiquitous phenomena & 1 \\
10 UCH{\sc ii}s & 50\% bipolarity, low collimation, $\dot{M}$ vs $L_{\rm{bol}}$ relation & 2 \\
3 YSOs & $\dot{M}$ vs $L_{\rm{bol}}$ still holds & 3 \\
11 YSOs & low collimation, no Hubble law, $\dot{M}$ vs $L_{\rm{bol}}$ not correlated & 4\\
69 HMPOs & 55\% bipolarity, accretion-outflow process suggested & 5 \\
26 HMPOs & 80\% bipolarity, no $\dot{M}$ vs $L_{\rm{bol}}$ but $M_{\rm{out}}$ vs $M_{\rm{core}}$, consistent & 6 \\
& with collimated outflows, similar to low-mass outflows & \\
139, all types & $M_{\rm{out}}$ \& $\dot{M}$ correlate with $L_{\rm{bol}}$, on average less collimated & 7, 8 \\
\hline
\multicolumn{3}{c}{High-spatial resolution interferometer and infrared studies}\\
\hline
G192.16, HCH{\sc ii} & wide-angle wind + molecular flow \& disk & 9,10,11 \\
IRAS20126, HCH{\sc ii} & precessing jet with disk & 12,13 \\
IRAS05358, HMPO & consistent with low-mass jets & 14 \\
W75N, HC/UCH{\sc ii} & energetics different to low-mass, not scaled up outflows & 15 \\
IRAS19410, HMPO & consistent with low-mass outflows & 16 \\
G35.2, HCH{\sc ii} & a cluster of collimated outflows & 17 \\
2 HMPOs & accretion based outflows & 18\\
4 HMPOs/HCH{\sc ii}s & same physical driving mechanisms as low-mass outflows & 19 \\
IRAS16547, HMPO & collimated jet, similar to low-mass jets & 20,21 \\
IRAS18151, HMPO & scaled-up versions of low-mass jets driven by disk accretion & 22 \\
G5.89, UCH{\sc ii} & multiple wide-angle outflows, harbors O5 star & 23 \\
\hline
\end{tabular}
\footnotesize{ The listed results just refer to mm studies of thermal
lines and infrared studies. Ref:
(1) \cite{shepherd1996a}, (2) \cite{shepherd1996b}, (3)
\cite{henning2000}, (4) \cite{ridge2001}, (5) \cite{zhang2001}, (6)
\cite{beuther2002b}, (7) \cite{wu2004}, (8) \cite{wu2005}, (9)
\cite{shepherd1998}, (10) \cite{devine1999}, (11) \cite{shepherd2001},
(12) \cite{cesaroni1999},
(13) \cite{shepherd2000},
(14) \cite{beuther2002d},
(15) \cite{shepherd2003}, 
(16) \cite{beuther2003a}, 
(17) \cite{gibb2003}, 
(18) \cite{su2004}, 
(19) \cite{beuther2004d}, 
(20) \cite{garay2003},
(21) \cite{brooks2003},
(22) \cite{davis2004}, 
(23) \cite{sollins2004b}}
\end{table}

The current situation is that different observations find results
which appear to contradict each other.  In order to resolve this
conflict, it is necessary to reevaluate the observations.  In
principle, three possibilities exist to solve the discrepancies: (a)
some observations are wrongly interpreted, (b) there exist physically
different modes of massive molecular outflows and thus massive star
formation, or (c) the observed sources are not directly comparable,
e.g., there might be evolutionary differences between the observed
sources which could translate into various evolutionary signatures of
the massive outflows.

It is possible that (a) applies to some of the single-dish results,
e.g., the initial claim of low collimation for massive molecular
outflows can be misinterpreted by the low spatial resolution of the
observation, the generally larger distances to the sources and
uncertainties due to the unknown inclination angles of the outflows.
However, the results based on high-spatial-resolution observations are
more significant and cannot be discarded in this manner. It appears
that physically different modes and/or evolutionary differences must
exist to explain all the observations.  While we cannot exclude the
possibility that different modes of star formation and their
associated outflows exist, it seems unlikely given the link between
accretion and outflow that has been established up to early B spectral
types.  Thus, here we propose an evolutionary scenario for massive
outflows which is capable of explaining the morphological and
energetic results within a consistent picture of massive protostellar
evolution.  This evolutionary scenario is based on the idea that stars
of all masses form via similar accretion-based processes that are
proposed for low-mass star formation (e.g., \cite{mckee2003}).

\subsection{A potential evolutionary scenario}

Reevaluation of the high-spatial resolution literature presented in
Table \ref{summary} shows that most of the youngest early B protostars
(HMPOs and those producing hypercompact H{\sc ii} regions) support the
idea that these high-mass outflows are driven by similar physical
processes as those seen in low-mass outflows. Contrary to this, three
sources in particular show different observational characteristics:
G5.89: which harbors an O5 star and multiple wide-angle outflows, and
G192.16 and W75N: two early B stars with UCH{\sc ii} regions. G192.16
and W75N VLA2 both appear to be relatively old (a few $\times 10^5$
years based on the CO dynamical timescale), and they have had adequate
time to reach the main sequence and produce significant ionizing
radiation.  The outflows are poorly collimated within 100 AU of the
central star and their outflow morphology on larger scales are
consistent with wind-blown bubbles.  Furthermore, it should be noted
that the single-dish outflow studies, which report a lower degree of
collimation for massive outflows preferentially observed sources with
UCH{\sc ii} regions (\cite{shepherd1996b,ridge2001}). In contrast,
most other single-dish surveys were directed toward samples of younger
sources, namely HMPOs, observing higher collimation degrees consistent
with low-mass outflows.

\begin{figure}[htb] \begin{center}
\includegraphics[width=9cm]{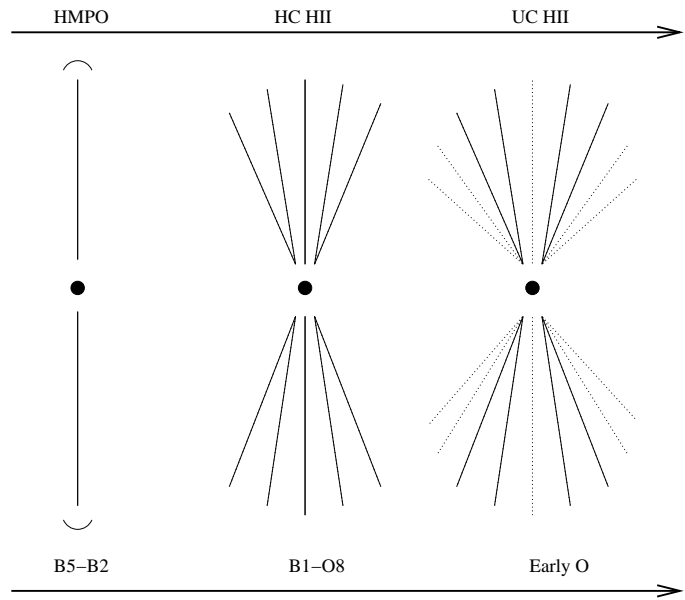} \end{center}
\caption{\footnotesize Sketch of the proposed evolutionary outflow
scenario.  The three outflow morphologies can be caused by two
evolutionary sequences: (top) the evolution of an early B-type star
from an HMPO via a HCH{\sc ii} region to an UCH{\sc ii} region, and
(bottom) the evolution of an early O-type star which goes through B-
and late O-type stages (only approximate labels) before reaching its
final mass and stellar luminosity.  }  \label{sketch}
\end{figure}

To date, the different observations have been interpreted either in
favor of: (1) similar driving processes for outflows of all masses
(e.g., \cite{beuther2002b,davis2004}); or (2) that massive stars form
via different physical processes like explosive coalescence or that
the outflows are powered by a combination of accretion and deflected
outflow.  (e.g., \cite{churchwell1999,bally2002}).  Here, we propose a
different, less controversial interpretation of the data: an
evolutionary sequence for massive protostars.  We consider two
possible evolutionary sequences which could result in similar
observable outflow signatures as shown in Figure \ref{sketch}.

\underline{\it (a) The outflow evolution of an early B-type star:} In
the accretion scenario, a B star forms via accretion through a
disk. During its earliest HMPO phase no HCH{\sc ii} region has formed
yet, and the disk outflow interaction produces collimated jet-like
outflows (Fig. \ref{sketch}\,(left), e.g., IRAS05358). At some point,
a HCH{\sc ii} forms and the wind from the central massive star
produces an additional less collimated outflow component. At that
stage the disk is not entirely destroyed yet, and jet and wind can
co-exist (Fig. \ref{sketch}\,(center), e.g., IRAS\,20126).  Evolving
further, a typical ultracompact H{\sc ii} region forms above and below
the accretion disk and the massive wind begins to dominate the whole
system (Fig. \ref{sketch}\,(right), e.g., W75N).  Remnants of the
initial jet might still exist, and density gradients within the
maternal core are likely to maintain at least a small degree of
collimation at that stage.

\underline{\it (b) The outflow evolution of an early O-type star:} An
intriguing aspect of the outflow studies is that so far extremely
collimated jet-like outflows have not been observed toward sources
earlier than B1.  It is possible that this is simply a selection
effect because the evolutionary timescale of early O stars is expected
to be only around $10^5$ years (\cite{mckee2002}).  Given the scarcity
of O stars and the short formation time, it is difficult to find
sources while they are still in the earliest stages of formation.
However, to form massive stars via accretion, the protostellar objects
must accrete even after the central object has reached the main
sequence.  In this scenario, at the beginning, the massive protostar
reaches a mid-B protostar-like state where it drives collimated
outflows (Fig. \ref{sketch}\,(left)).  As the object continues to
accrete it becomes an early-B to late-O star, forming a HCH{\sc ii}
region while the wind and jet components could co-exist
(Fig. \ref{sketch}\,(center)).  Finally, as the central star evolves,
it may eventually become a mid- to early-O star.  The increased
radiation from the central star would generate significant
Lyman-$\alpha$ photons and ionize the molecular outflowing gas even at
large radii.  The result may be an increase in the plasma pressure at
the base of the outflow which could overwhelm the collimating effects
of a magnetic field (e.g., \cite{shepherd2003b,
koenigl2000}). Incorporating the additional physical effects of
high-mass protostars on the surrounding disk and envelope is one of
the challenges of theoretical massive outflow modeling (e.g.,
\cite{konigl1999}).  In this scenario, it would be intrinsically
impossible to ever observe collimated jet-like outflows from very
young early O-type (proto)stars.

\subsection{Discussion and potential caveats}

This scenario is not the only possible interpretation, and different
physical processes could take place at the earliest stages of massive
star formation.  This reevaluation is based only on imaging data of
thermal lines in the mm and infrared bands.  However, many studies of
outflows have also been undertaken by the means of maser studies,
especially H$_2$O maser.  While some maser studies found H$_2$O maser
emission consistent with jet-like outflows (e.g. \cite{kylafis1999}),
other studies find shell-like expanding features which cannot be
explained by collimated jets (e.g.,
\cite{patel2000,torrelles2001,torrelles2003}). The most intriguing
case is the H$_2$O maser study in W75N: Torrelles et al.~(2003) find
on projected spatial scales of 1400\,AU that the two cm continuum
sources VLA1 and VLA2 drive very different types of outflows, the
first one appears to be a collimated jet whereas the second one rather
resembles a wind-like shell.  As the two sources are part of the same
core, environmental properties cannot explain the difference.  Both
sources appear to have a similar luminosity (B1-B2 spectral types)
although the spectral type derived for VLA1 should be considered an
upper limit due to likely strong contamination by ionizing flux
produced by shock waves in the jet (\cite{shepherd2004}).  Assuming
the VLA1 jet is produced by an early B star, then the difference in
morphology would not be a question of the mass of the central source
as well.  Thus, Torrelles et al.~(2003) propose that an evolutionary
explanation is most likely.  They cannot determine which source is
younger but based on the extent of the maser emission -- the
non-collimated features are more compact -- they suggest that the
collimated jet might belong to the more evolved source.  This would be
a counter-example to the scenario presented in this paper as well as
to the scenarios discussed for low-mass outflows (e.g.,
\cite{andre2000}).  However, maser emission appears to be very
selective, and many known outflow sources have no H$_2$O maser
emission at all (e.g., \cite{beuther2002b}).  Therefore, it is also
possible that the collimated part of the outflow from VLA2 is simply
not depicted by the data, and the non-collimated features have been
excited more recently by the central HCH{\sc ii} region. Furthermore,
the cm continuum emission from VLA1 may be due primarily to the jet
(\cite{shepherd2004}).  Thus, it is feasible that the central source
of VLA1 is either less massive than VLA2 or it is younger and has not
formed a HCH{\sc ii} region yet.  Therefore, we might only observe a
collimated feature toward this source.  The latter scenario is
consistent with the evolutionary sequence presented here.  To properly
differentiate between both scenarios in this source requires that we
find other means to determine the age of the powering sources VLA1 and
VLA2.

It should be stressed that the proposed evolutionary scenario is
still only a potential, qualitative explanation for the observed
outflow features which must be tested against both theory and
observations.
\bibliographystyle{kapalike}

\input{bibliography.bbl}
\end{document}

%% file: refs.tex
\def\aj{AJ}%
          % Astronomical Journal
\def\araa{ARA\&A}%
          % Annual Review of Astron and Astrophys
\def\apj{ApJ}%
          % Astrophysical Journal
\def\apjl{ApJ}%
          % Astrophysical Journal, Letters
\def\apjs{ApJS}%
          % Astrophysical Journal, Supplement
\def\ao{Appl.~Opt.}%
          % Applied Optics
\def\apss{Ap\&SS}%
          % Astrophysics and Space Science
\def\aap{A\&A}%
          % Astronomy and Astrophysics
\def\aapr{A\&A~Rev.}%
          % Astronomy and Astrophysics Reviews
\def\aaps{A\&AS}%
          % Astronomy and Astrophysics, Supplement
\def\azh{AZh}%
          % Astronomicheskii Zhurnal
\def\baas{BAAS}%
          % Bulletin of the AAS
\def\jrasc{JRASC}%
          % Journal of the RAS of Canada
\def\memras{MmRAS}%
          % Memoirs of the RAS
\def\mnras{MNRAS}%
          % Monthly Notices of the RAS
\def\pra{Phys.~Rev.~A}%
          % Physical Review A: General Physics
\def\prb{Phys.~Rev.~B}%
          % Physical Review B: Solid State
\def\prc{Phys.~Rev.~C}%
          % Physical Review C
\def\prd{Phys.~Rev.~D}%
          % Physical Review D
\def\pre{Phys.~Rev.~E}%
          % Physical Review E
\def\prl{Phys.~Rev.~Lett.}%
          % Physical Review Letters
\def\pasp{PASP}%
          % Publications of the ASP
\def\pasj{PASJ}%
          % Publications of the ASJ
\def\qjras{QJRAS}%
          % Quarterly Journal of the RAS
\def\skytel{S\&T}%
          % Sky and Telescope
\def\solphys{Sol.~Phys.}%
          % Solar Physics
\def\sovast{Soviet~Ast.}%
          % Soviet Astronomy
\def\ssr{Space~Sci.~Rev.}%
          % Space Science Reviews
\def\zap{ZAp}%
          % Zeitschrift fuer Astrophysik
\def\nat{Nature}%
          % Nature
\def\iaucirc{IAU~Circ.}%
          % IAU Cirulars
\def\aplett{Astrophys.~Lett.}%
          % Astrophysics Letters
\def\apspr{Astrophys.~Space~Phys.~Res.}%
          % Astrophysics Space Physics Research
\def\bain{Bull.~Astron.~Inst.~Netherlands}%
          % Bulletin Astronomical Institute of the Netherlands
\def\fcp{Fund.~Cosmic~Phys.}%
          % Fundamental Cosmic Physics
\def\gca{Geochim.~Cosmochim.~Acta}%
          % Geochimica Cosmochimica Acta
\def\grl{Geophys.~Res.~Lett.}%
          % Geophysics Research Letters
\def\jcp{J.~Chem.~Phys.}%
          % Journal of Chemical Physics
\def\jgr{J.~Geophys.~Res.}%
          % Journal of Geophysics Research
\def\jqsrt{J.~Quant.~Spec.~Radiat.~Transf.}%
          % Journal of Quantitiative Spectroscopy and Radiative Trasfer
\def\memsai{Mem.~Soc.~Astron.~Italiana}%
          % Mem. Societa Astronomica Italiana
\def\nphysa{Nucl.~Phys.~A}%
          % Nuclear Physics A
\def\physrep{Phys.~Rep.}%
          % Physics Reports
\def\physscr{Phys.~Scr}%
          % Physica Scripta
\def\planss{Planet.~Space~Sci.}%
          % Planetary Space Science
\def\procspie{Proc.~SPIE}%
          % Proceedings of the SPIE
\let\astap=\aap
\let\apjlett=\apjl
\let\apjsupp=\apjs
\let\applopt=\ao